\documentclass[12pt]{iopart}

\usepackage{iopams}
\usepackage{graphicx}
\usepackage{subfigure}
\usepackage{color}
\usepackage{latexsym}
\usepackage{array}

\begin{document}

\title[]{Run-and-Tumble-Like Motion of Active Colloids in Viscoelastic Media}

\author{Celia Lozano$^{1,2,*}$, Juan Ruben Gomez-Solano$^{1,*}$, and Clemens
Bechinger$^{1,2,*}$}

\address{$^1$2. Physikalisches Institut, Universit\"at Stuttgart,
Pfaffenwaldring 57, 70569 Stuttgart, Germany}
\address{$^2$Max-Planck-Institute for Intelligent Systems,
Heisenbergstrasse 3, 70569 Stuttgart, Germany}

\address{*(present address) Fachbereich Physik, Universit\"at Konstanz, Konstanz  D-78457, Germany}

\ead{clemens.bechinger@uni-konstanz.de}

\vspace{10pt}

\begin{abstract}

Run-and-tumble (RNT) motion is a prominent locomotion strategy employed by many living microorganisms. It is characterized by straight swimming intervals (runs), which are interrupted by \emph{sudden} reorientation events (tumbles). In contrast, directional changes of synthetic microswimmers (active particles, APs) are caused by rotational diffusion, which is superimposed with their translational motion and thus leads to rather \emph{continuous} and slow particle reorientations. Here we  demonstrate that active particles can also perform a swimming motion where translational and orientational changes are disentangled, similar to RNT. In our system, such motion is realized by a viscoelastic solvent and a periodic modulation of the self-propulsion velocity. Experimentally, this is achieved using light-activated Janus colloids, which are illuminated by a time-dependent laser field. We observe a strong enhancement of the effective translational and rotational motion when the modulation time is comparable to the relaxation time of the viscoelastic fluid. Our findings are explained by the relaxation of the elastic stress, which builds up during the self-propulsion, and is suddenly released when the activity is turned off. In addition to a better understanding of active motion in viscoelastic surroundings, our results may suggest novel steering strategies for synthetic microswimmers in complex environments.

\end{abstract}


\section{Introduction}

Synthetic microswimmers or active particles (APs) \cite{Paxton,BechingerReview,Dreyfus,howse,palacci,jiang,buttinoni,romanczuk,zoettl} currently receive considerable attention because their properties resemble the behavior of living microorganisms \cite{Cates2012,elgeti2015,Schwarz,Polin}. Similar to their biological counterparts, APs can form clusters \cite{Theurkauff2012,Palacci2013,Buttinoni2013}, respond to external chemical \cite{Liebchen} and light \cite{Lozano2016} gradients or to gravity \cite{Borge2014} and external flows \cite{Palacci2016}. In both cases, the translational mean square displacement (MSD) displays a short-time ballistic and a long-time effective diffusive behavior. Despite such common features between biological and synthetic microswimmers, notable differences occur. For instance, the trajectories of \textit{E. Coli} are comprised of alternate periods of rather straight segments with almost constant velocity $v$ (\textit{runs}), and abrupt short-lived reorientation events (\textit{tumbles}) which result in rather sudden directional changes \cite{BergRT,Rupprecht}. 
In contrast, the angular dynamics of synthetic microswimmers is determined by their rotational diffusion which - for micron-sized particles - is on the order of several tens up to hundreds of seconds \cite{catesRT,Schwarz}. Because orientational and positional changes are superimposed, directional changes of active particles are typically rather smooth compared to their biological counterparts.

Here we  demonstrate that APs can also perform a swimming motion where translational and orientational changes are decoupled in time. Such run-and-tumble-like propulsion scheme can be induced in a suspension of APs being immersed in a viscoelastic medium when their self-propulsion is periodically turned on and off with modulation time $T_{mod}$. While the particles perform a rather straight forward swimming motion during the on-period, they experience a pronounced elastic recoil when the propulsion is turned off. Similar recoils have been previously observed in microrheological experiments where a (passive) colloidal particle was externally driven by an optical laser tweezer through a viscoelastic liquid ~\cite{RubenNJP}. When the optical trap was suddenly turned off, the relaxing microstructure of the fluid caused a transient backward motion of the particle, i.e. an elastic recoil. In case of an AP with time-modulated self-propulsion as studied here, such recoils are found to strongly affect both, its translational and orientational dynamics. We observe that the shape of the trajectories strongly depends on $T_{mod}$ and that particle reorientation is most rapid when it is close to the relaxation time $\tau$ of the fluid. Our experimental findings suggest novel steering strategies for active colloids in e.g. crowded systems \cite{Li} or within patterned confinements \cite{Yazdi}.

\begin{figure}
	\includegraphics[width=0.9\columnwidth]{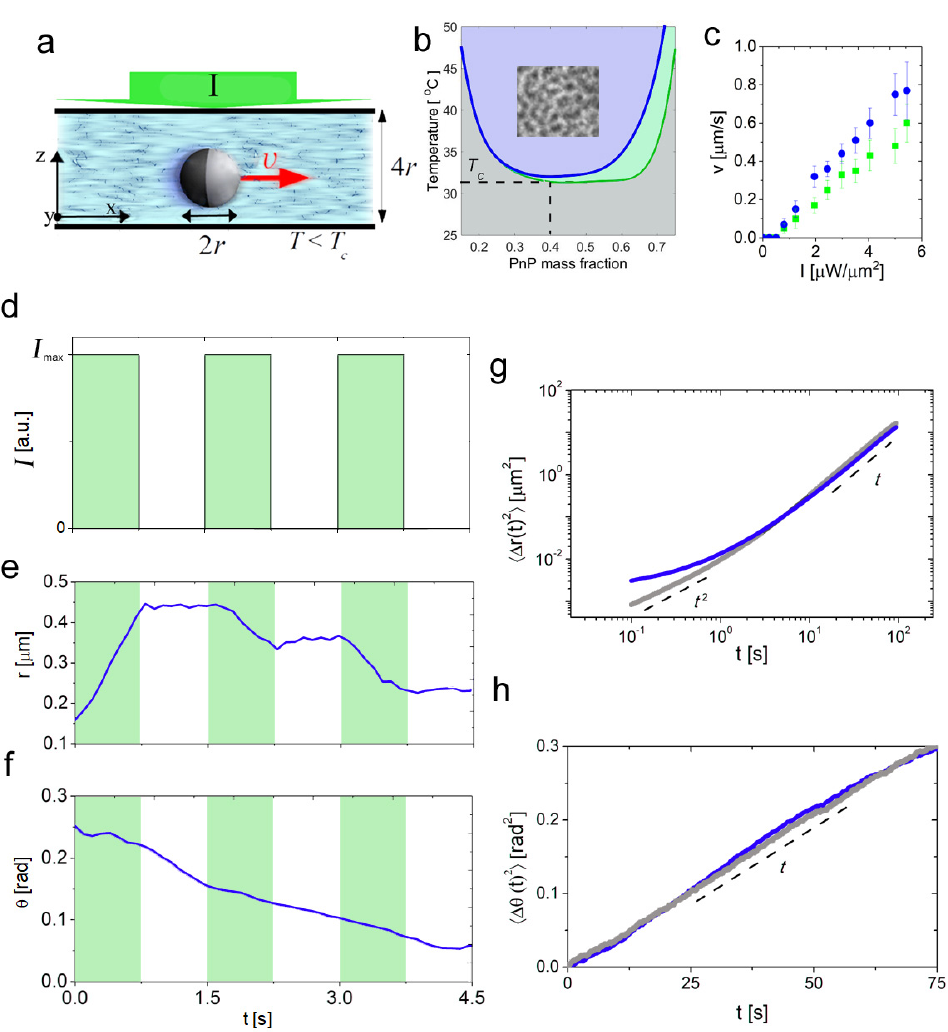}
\caption{ (a) Schematic illustration of the sample cell containing a dilute suspension of light-activated Janus particles in a fluid.  (b) Phase diagram of a Newtonian binary critical mixture of water and PnP (blue) with the critical point at $T_c$ = $31.4^{\circ}$C and 0.4 PnP mass fraction. The green line corresponds to the phase diagram after addition of $0.05\%$ polyacrylamide (PAAm), which renders the fluid viscoelastic. Inset: snapshot of the spinodal demixing of the viscoelastic fluid at $T$ = $32^{\circ}$C. The nature of the demixing process is not affected by the presence of PAAm. (c) Self-propulsion velocity of a Janus particle with $2R = 7.75$~$\mu$m as a function of the light intensity under steady illumination conditions for Newtonian  (blue circles) and viscoelastic mixture (green squares). (d) Intensity  protocol to realize a periodic modulation of particle self-propulsion. (e),(f) Positional and orientational configuration of an active particle whose self-propulsion is modulated in time with $T_{mod}= 1.5s$ and $I_{max}=4$~$\mu$W/$\mu$m$^2$. (g),(h) Translational and angular mean-square displacements for periodical modulation (same conditions as in (e,f)) (blue) and homogeneous illumination with $I_{max}=2$~$\mu$W/$\mu$m$^2$ to ensure identical conditions (gray).}
\label{fig:fig1}
\end{figure}

\section{Experimental description}

\begin{figure}
	\includegraphics[width=\columnwidth]{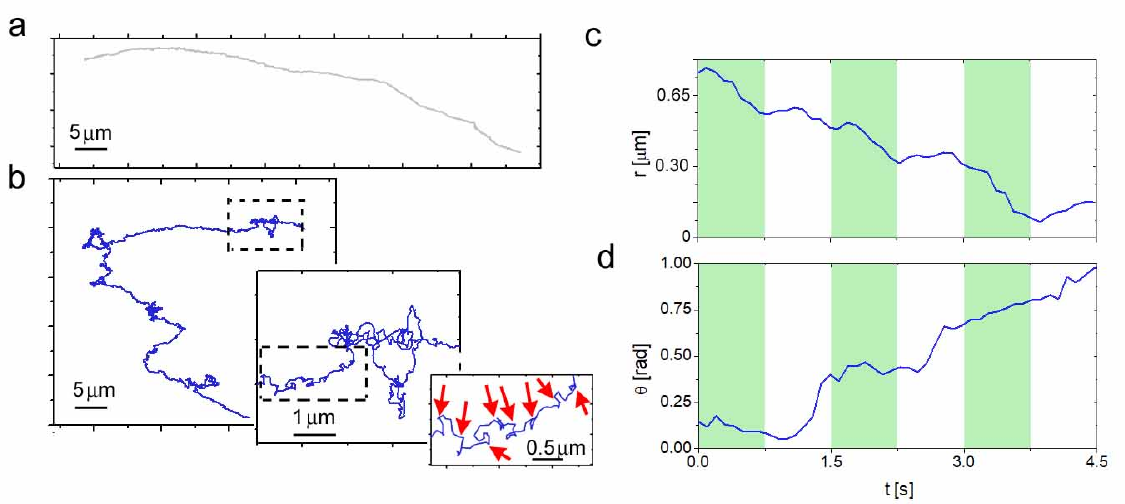}
\caption{ (a) Typical trajectory of a self-propelled particle in a viscoelastic fluid at constant illumination with $I=2$~$\mu$W/$\mu$m$^2$. (b) Trajectory of the same particle for $T_{mod}= 1.5s$ and $I=4$~$\mu$W/$\mu$m$^2$. The arrows indicate events where the self-propulsion is turned off. (c),(d) Positional and orientational configuration of an active particle with periodically modulated self-propulsion for the same conditions as in (b).}
\label{fig:fig2}
\end{figure}

As active colloids, we used spherical silica particles with diameter $2R = 7.75$~$\mu$m, which are half-coated by a 50 nm sputtered carbon layer (Fig.1a). Such Janus-particles are immersed in a critical mixture of water and propyleneglycol n-propyl ether (PnP) with 0.4 PNP mass fraction. To render the fluid viscoelastic, we added $0.05\%$ polyacrylamide (PAAm) ($M_w=18 \times 10^{6}$). The solution exhibits a lower critical point as shown by the green line in Fig.1b. In the semidilute regime, such solutions are viscoelastic with a stress-relaxation time $\tau=1.50 \pm  0.10$s (for details, see~\cite{RubenPRL}). As experimentally confirmed, the solution displays shear thinning behavior for Weissenberg numbers $\mathrm{Wi} > 1$, where Wi$=v\tau/2R$, with $v$ the particle propulsion velocity \cite{gomezsolano1}. It should be emphasized that in all experiments shown below, $\mathrm{Wi} < 0.3$ where the fluid has a constant (zero) shear viscosity $\eta_0$=$0.100 \pm 0.015$ Pa s. In this regime, the mean viscous friction force on a particle has been confirmed to be proportional to its velocity ~\cite{RubenNJP}. 

A small amount of APs is immersed in the mixture and contained in a thin sample cell with height $h$ $\approx 4R $ (Fig.1a). Under such spatial confinement translational and rotational dynamics is limited to two dimensions ~\cite{das,popescu,rubenSPP}. In the absence of self-propulsion, the translational and rotational diffusion coefficients determined from our experiments are $D^0_t = (3.69 \pm 0.60)\times 10^{-4}\,\mu \mathrm{m}^2\,\mathrm{s}^{-1}$ and $D^0_r = (1.84 \pm 0.40)\times 10^{-5}\,\mathrm{s}^{-1}$, respectively. This value is about 50\% below the corresponding Stokes-Einstein values and caused by hydrodynamic coupling to the walls \cite{brenner}. The sample cell is thermally coupled to a water bath which is kept constant below the critical temperature at $T=25 \pm 0.1^{\circ}$C. When the cell is illuminated with laser light ($\lambda$ = 532 nm), it is absorbed by the carbon caps, which results in local demixing of the fluid (for our intensities this demixing region is at least one order of magnitude smaller than the particle size \cite{rubenSPP}). This leads to a self-diffusiophoretic motion whose propulsion velocity is controlled by the incident light intensity as shown by the green symbols in Fig.1b \cite{buttinoni,samin,rubenSPP}. To rule out a possible temperature dependence of the rheological properties of the viscoelastic mixture, which would influence the viscous drag experienced by APs, we have measured the shear-rate dependent viscosity of our system between $25^{\circ}$C (bath temperature) and $31^{\circ}$C, the latter being close to $T_C$ (for further details see appendix A). Within our experimental resolution, only a negligible temperature-dependence is observed.

Time-dependent propulsion forces were created by periodic variation of the laser intensity with $T_{mod}$ by a mechanical shutter, which periodically switches the illumination intensity between $I_{max}$ and zero (Fig. 1d). In our protocol, we have chosen identical \emph{on} and \emph{off} duration times which guarantees that the time-averaged incident light intensity is independent of $T_{mod}$. Particle positons and orientations were obtained by digital video microscopy with a frame rate of 10 Hz and 50nm of spatial resolution. The particle orientation was determined directly from the optical contrast due to the carbon cap ~\cite{RubenPRL}.

\section{Experimental Results}

\begin{figure}
        \includegraphics[width=0.7\columnwidth]{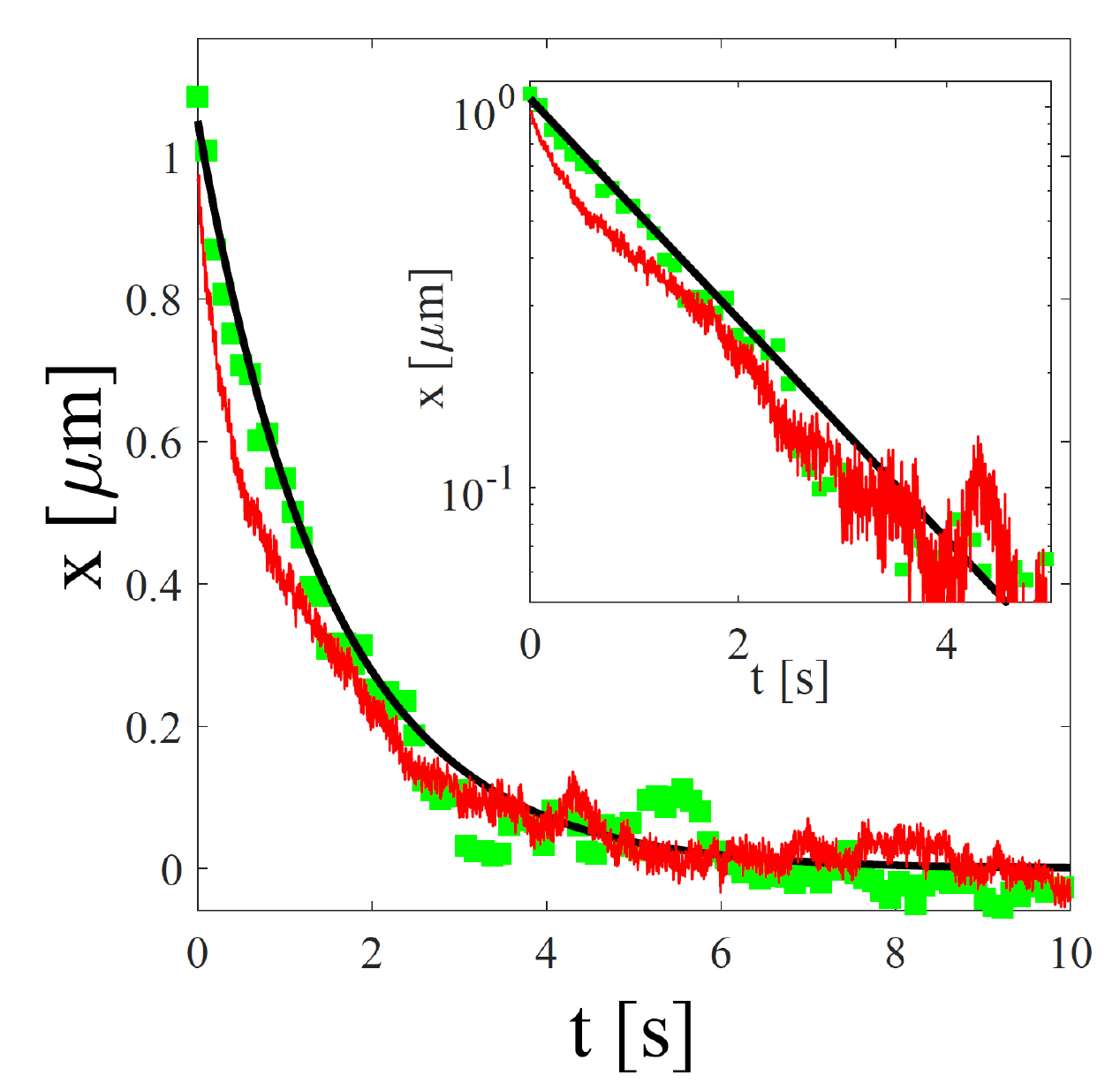}
\caption{Comparison of the recoil motion of an AP when the illumination is suddenly turned off (green) and that of a colloidal particle which is suddenly released from a moving optical trap (red). The particle velocity prior to the recoil (t=0) was in both cases $v \approx 0.4\mu m/s$ and the particle size $2R = 7.75 \mu m$. The solid black line corresponds to an exponential fit $x \propto \exp(-t/\tau)$, where $\tau = 1,5$~s. Inset: semilog representation of the main figure.}
\label{fig:Fig3}
\end{figure}

Before discussing the behavior of active particles in viscoelastic fluids, we will briefly discuss their motion in a Newtonian binary mixture, i.e. in absence of the PAAm. As shown by the blue lines in Fig.1b,c phase diagram of a water-PNP mixtures is little affected by the presence of PAAm and only leads to small changes in the critical temperature and the light-induced particle propulsion. Figs.1e,f show the in-plane positional and orientational component of an active particle which is subjected to a periodic laser illumination. In contrast to the translational motion, which alternates between self-propulsion (light on) and Brownian diffusion (light off), no differences in the angular component are found. The latter is expected, because for homogeneous illumination, the particle orientation is entirely determined by its rotational diffusion and independent of self-propulsion. To compare the particle's motion in Newtonian liquids for continuous and periodic illumination, we have analyzed the positional and orientational component of their trajectories. To allow for a direct comparison, the time-averaged illumination intensity was identical in both cases. Figs.1g,h show the positional and angular mean square displacements (MSDs) for periodic, $T_{mod}=1.5 s$ (blue), and steady (gray) illumination, respectively. Apart from deviations in the short time-behavior of the translational MSD \cite{hartmut}, the diffusive long-time behavior is identical for periodic and steady illumination. This demonstrates that a steady state in the particle motion is reached within the illumination period. Such behavior is consistent with the fact that the temperature profile around a micron-sized capped colloid suspended in a liquid with thermal diffusivity $\alpha \sim 10^{-7}m^2/s$ reaches its steady state after a time scale $R^2/\alpha \sim10^{-3}$s  \cite{rubenSPP}. Note that this is at least two orders of magnitude shorter than the time-scales of our experiments, namely $\tau$ = 1.5 s and the smallest modulation period $T_{mod}$ = 0.2 s in our experiments.

In the following, we discuss our observations after adding PAAm to the critical solution, i.e. in a viscoelastic fluid. Because the viscosity is enhanced by a factor of 25 after adding the polymer ($\eta_0$=$0.100 \pm 0.015$ Pa s), the particle's diffusive  reorientation dynamics becomes much slower compared to above. This is seen in Fig.2a, which shows the rather straight particle trajectory under constant illumination over 400s \cite{footnote1}. When the active motion is periodically turned on and off, drastic directional changes appear. These reorientation events mainly occur when the illumination is turned off, as  shown by the arrows in Fig.2b.
The effect of the modulated propulsion on the particle's trajectory is seen in more detail in Fig.2c. During each illumination period the particle is translated by several hundreds of nanometres. In the absence of illumination, a much smaller reverse motion occurs. Contrary to the Newtonian case, here pronounced differences between periods with and without illumination are also observed in the orientational particle dynamics (Fig.2d): while only slow orientational changes occur during self-propulsion, abrupt directional changes arise when the illumination is turned off.

The above observations can be explained by an elastic recoil, which is experienced by the particle after illumination is turned off. The presence of such recoils has been recently demonstrated in microrheological experiments, where a colloidal particle was driven by a moving optical tweezer through a viscoelastic solution \cite{RubenNJP}. When the driving force was suddenly removed, a back motion due to the recovery of the deformed fluid's microstructure was observed. Such behavior, which is absent in Newtonian liquids, is caused by the rather long (seconds) structural relaxation times being characteristic for viscoelastic fluids \cite{Bird}. Fig.3 (red curve) shows the result of such a recoil experiment in the same viscoelastic solvent and for the same bath temperature $T=25^{\circ}$C and particle size $2R = 7.75 \mu m$ as used in the present study. The velocity before release was set to $v \approx 0.4 \mu m/s$ which results in an exponential decay with a decay time of $\tau = 1.5s$ (black line) \cite{footnote2}. The green curve shows the transient of an equally sized active particle, which was originally self-propelling with velocity $v \approx 0.4 \mu m/s$ and whose active motion was suddenly stopped by turning the illumination off. In both cases, we observe the same exponential decay, which suggests the occurrence of elastic recoil in APs in viscoelastic fluids with modulated self-propulsion.

In analogy with the Deborah number (De), which is used in rheology to characterize flows of viscoelastic materials under time-dependent deformations, in the following we will quantify the motion of active particles with periodic propulsion modulation by

\begin{equation}\label{eq:De}
        De=\frac{\tau}{T_{mod}}.
\end{equation}

\begin{figure}
\includegraphics[width=\columnwidth]{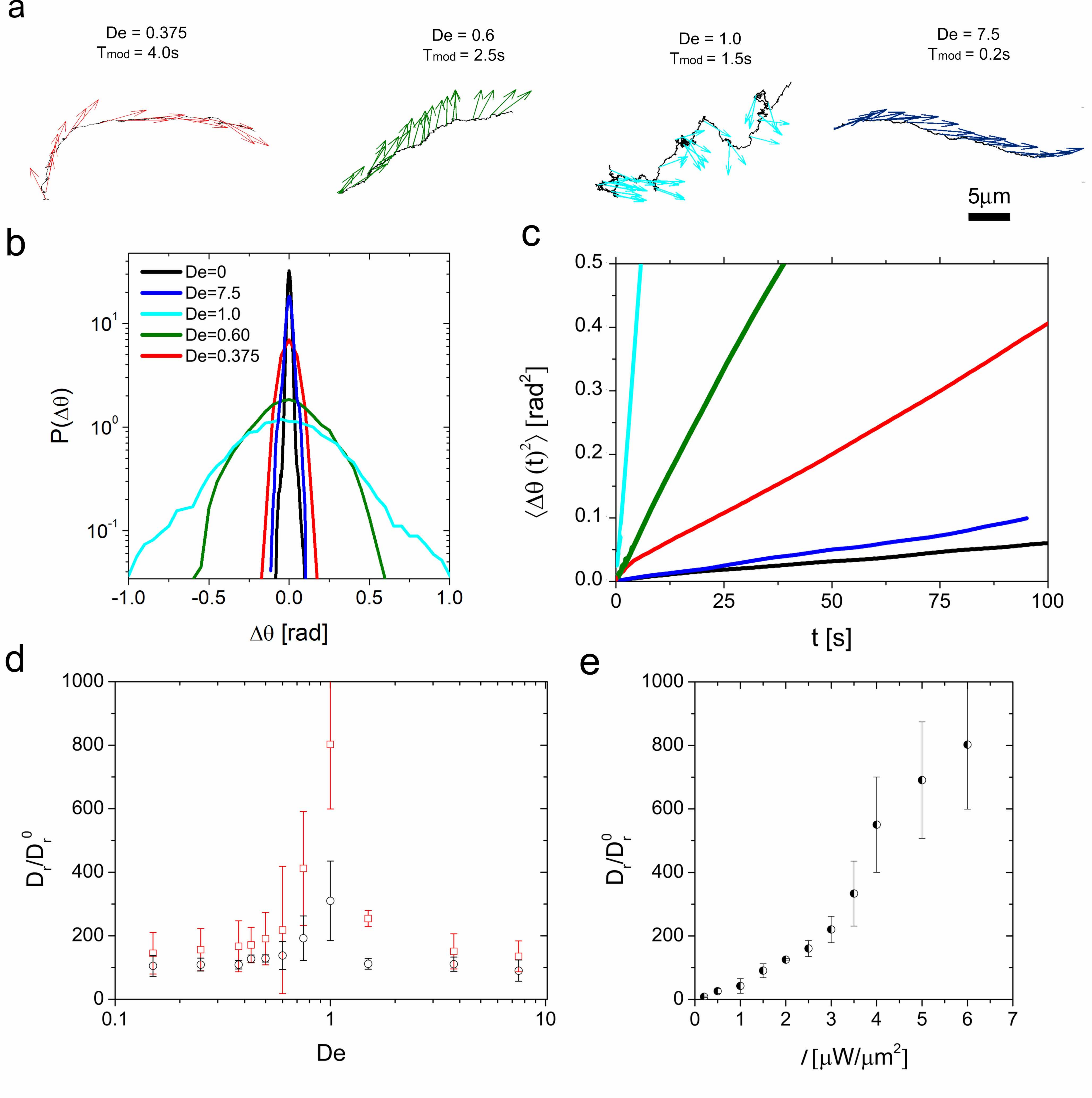}
\caption{(a) Particle trajectories (400 s each) for different modulation times and for $I_{max}=4$~$\mu$W/$\mu$m$^2$. The local particle orientations are indicated by arrows. (b) Normalized probability distribution functions of orientational changes $\Delta \theta$ during $\Delta t=T_{mod}$ for different De. (c) Corresponding orientational mean square displacements which indicate an effective diffusive behavior at long times. (d) Normalized orientational diffusion coefficient vs. De for $I_{max}= 2$~$\mu$W/$\mu$m$^2$ (black circles), $4$~$\mu$W/$\mu$m$^2$ (red squares). (e)  Normalized orientational diffusion coefficient vs. intensity $I$ at De=1.}
\label{fig:fig4}
\end{figure}

Fig.4a shows typical APs trajectories measured over 400 s duration and for $T_{mod}= 0.2, 1.5, 2.5, 4.0 s$ (corresponding to De = 7.5, 1, 0.6, 0.375) and for $I_{max}$=$4$~$\mu$W/$\mu$m$^2$ , where the arrows show the corresponding particle orientation determined from the video images by detecting the cap orientation. At low and high De, the trajectories are rather straight, but become quite wiggly at De =1. 

To quantify the change of the particle orientation $\Delta \theta$ during the off-period, we have calculated the normalized probability distributions of $\Delta \theta$ (Fig.4b). Compared to constant illumination (De=0), the distributions become wider, which suggests
that the particle reorientation is strongly modified by $T_{mod}$. Around De=1, the width of the distribution is the largest and exhibits orientational changes of several tens of degrees after a single modulation.

We have also calculated the mean-square displacement of the angular particle orientation $\theta$ , $\langle \Delta \theta (t)^2 \rangle = \langle [\theta(t_0 + t) - \theta(t_0)]^2  \rangle $, where the brackets denote time averaging over $t_0$. We plot the behavior of $\langle \Delta \theta (t)^2 \rangle$ for different De in Fig.4c.  Towards long times, the growth of $\langle \Delta \theta (t)^2 \rangle $ becomes linear with $t$. Interestingly, the slope  changes with De. Therefore, in the long-time limit we can determine an effective rotational diffusion coefficient $D^{(\mathrm{De})}_r$, which depends on the Deborah number, as

\begin{equation}\label{eq:effDr}
	\langle \Delta \theta (t)^2 \rangle  = 2D^{(\mathrm{De})}_r t.
\end{equation}

From the slope we obtain a De-dependent $D_r$ which is plotted (normalized to $D_r^0$) in Fig.4d. The graph displays a maximum at De=1 with an enhancement factor around 800. 
We also investigated the dependence of $D_r/D_r^0$ on the propulsion velocity. The data, which are plotted in Fig.4e for De=1, show a remarkably strong and nonlinear increase of the rotational diffusion on $I_{max}$. Qualitatively, this can be understood by considering that the elastic stress in the fluid increases when the particle is faster. Accordingly, the deformation of the fluid and, thus, the recoil (being responsible for the particle's reorientation dynamics) becomes enhanced.

As mentioned above, the elastic recoil also affects the translational particle dynamics. This is shown in Fig.5a, where we show that the translational MSD $\langle \Delta r (t)^2 \rangle$, (calculated in the same manner as the angular MSD) obtained from trajectories of at least 3600 s for different De.  The deviations from a linear behavior (oscillations) at short times are due to the propulsion modulation. Similar to the situation of Newtonian liquids (Fig.1g), the long-time behavior is linear in time, i.e. entirely diffusive ($\langle \Delta r (t)^2 \rangle$ = $4D_t t$ ). 

Finally, we also have investigated the dependence of $D_t/D_t^0$ vs. De for $I_{max}= 4$~$\mu$W/$\mu$m$^2$ (red squares) and $2$~$\mu$W/$\mu$m$^2$ (black circles).
Independent of the illumination intensity, we observe a strong enhancement of $D_t/D_t^0$  around De=1. A dependence of the translational diffusion coefficient for APs with time modulated propulsion force has been theoretically predicted also for viscous fluids \cite{hartmut}. The magnitude of the enhancement observed by us is, however, considerably larger and, obviously, results from the periodic elastic recoils which are experienced by the AP when its self-propulsion is modulated in time.

\begin{figure}
	\includegraphics[width=0.6\columnwidth]{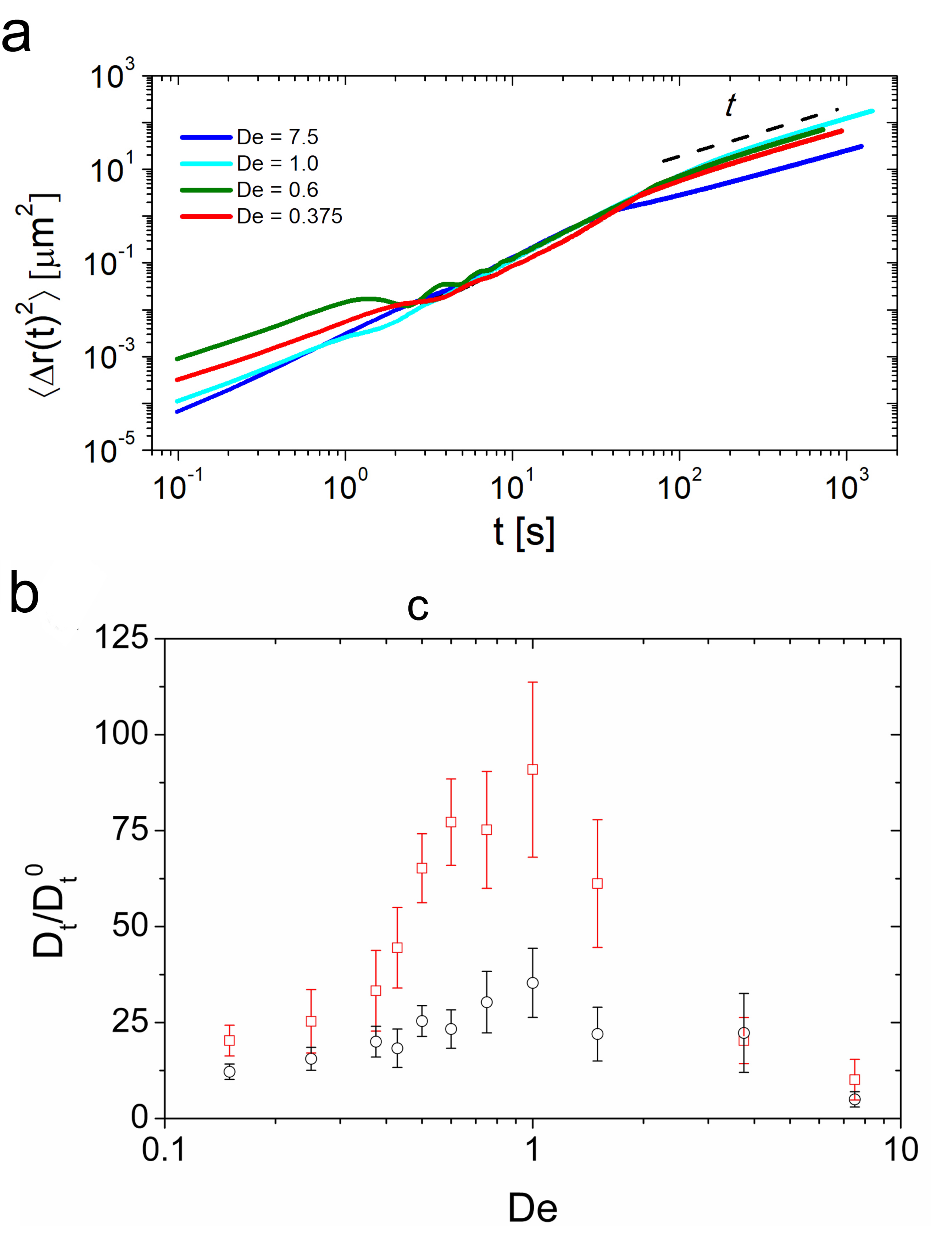}
\caption{(a) Translational mean square displacement for different  De = 7.5, 1.5, 1, 0.6, 0.375, for  $I_{max}=4.0$ $\mu$W$/$$\mu$$m^2$. The solid line with slope equal to 1 shows the long-time translational diffusivity. (b) Normalized translational diffusion coefficient vs. De for $I_{max}= 2$~$\mu$W/$\mu$m$^2$ (black circles) and $4$~$\mu$W/$\mu$m$^2$ (red squares). }
\label{fig:fig5}
\end{figure}

\section{Conclusions}
In contrast to Newtonian fluids, where orientational changes of APs are usually determined by their slow rotational diffusion, we have shown that much faster directional changes can be induced in viscoelastic environments when the self-propulsion of APs is periodically modulated in time. As a result of the modulated propulsion strength, the translational and directional dynamics becomes disentangled in time. Our observations are explained by the elastic stress, which builds up during self-propulsion and which is suddenly released when the propulsion is turned off. Because this relaxation process couples to the particle dynamics, it results in an increased translational and rotational particle diffusion, which peaks when modulation time is identical to the microstructural relaxation time of the fluid. In addition to a variation of the modulation time, we expect that the particle's dynamics should also depend on the functional shape of the modulation (rectangular, sinusoidal, sawtooth, etc.) and thus allows to optimize the locomotion to different types of environments. Because the particle reorientation dynamics is key for the stability of clusters and the motion of APs through topographical landscapes, the possibility to control the rotational motion of APs may suggest novel strategies for particle steering under  conditions where rapid directional changes are required.

\section*{Acknowledgments}
We thank H-J. K\"ummerer, C. Mayer and U. Rau for their technical support. C. B. acknowledges financial support from the German Research Foundation (DFG) through the priority programme SPP 1726 on microswimmers and by the ERC Advanced Grant ASCIR (Grant No.693683). J-R.G-S. was supported by Deutsche Forschungsgemeinschaft (DFG) grant No.  GO 2797/1-1.

\appendix

\section{Rheological properties of the polymer critical mixture}\label{appa}
We have characterized the temperature-dependence of the viscosity $\eta$ of the viscoelastic mixture used in our investigation as a function of the applied shear rate $\dot{\gamma}$ (brand name). Fig.A1 shows the results for 
two different temperatures:  $T = 25^{\circ}$C (dashed green), which corresponds to the bath temperature of our solvent and 
$T = 31^{\circ}$C (solid red) which is close to the critical point of the mixture and thus near the temperature of the illuminated cap. Within our experimental resolution, we do not find a temperature dependence within this range.
This implies that no gradients in the viscosity around an anisotropically heated Janus particle are present within this temperature range.

\begin{figure}
        \includegraphics[width=0.6\columnwidth]{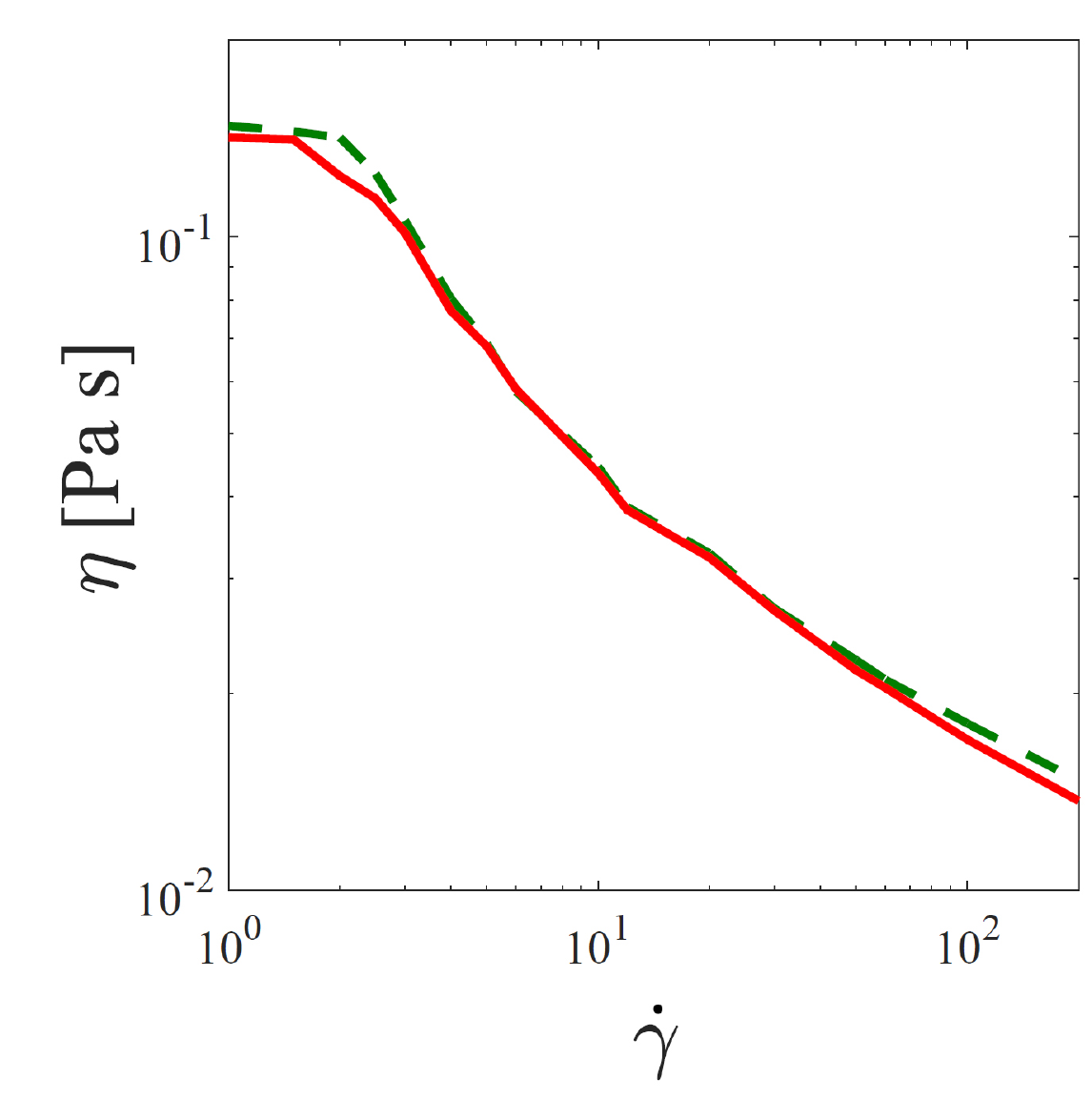}
\caption{ Viscosity $\eta$ as a function of the shear rate $\dot{\gamma}$ for the viscoelastic PAAm critical mixture measured at three different temperatures:  $T = 25^{\circ}$C (dashed green line), and $T = 31^{\circ}$C (solid red line).}
\label{fig:figapp}
\end{figure}


\section*{References}


\begin{thebibliography}{9}


\bibitem{Paxton}
W. F. Paxton, K. C. Kistler, C. C. Olmeda, A. Sen, S. K. St Angelo, Y. Cao, T. E. Mallouk, P. E. Lammert, V. H. Crespi. J Am Chem Soc. {\bf126}(41):13424-31 (2004).


\bibitem{BechingerReview}
 C. Bechinger, R. Di Leonardo, H. Löwen, C. Reichhardt, G. Volpe, G. Volpe.. J Am Chem Soc. {\bf88}, 045006 (2016).

 \bibitem{Dreyfus}
R. Dreyfus, J. Baudry, M. L. Roper, M. Fermigier, H. A. Stone, J. Bibette. Nature {\bf437}, 862-865 (2005).

\bibitem{howse}
J. R. Howse, R. A. L. Jones, A. J. Ryan, T. Gough, R. Vafabakhsh, and R. Golestanian, Phys. Rev. Lett. {\bf 99}, 048102 (2007).

\bibitem{palacci}
J. Palacci, C. Cottin-Bizonne, C. Ybert, and L. Bocquet, Phys. Rev. Lett. {\bf 105}, 088304 (2010).

\bibitem{jiang}
H.-R. Jiang, N. Yoshinaga, and M. Sano, Phys. Rev. Lett. {\bf105}, 268302 (2010).

\bibitem{buttinoni}
I. Buttinoni, G. Volpe, F. K\"ummel, G. Volpe, and C. Bechinger, J. Phys.: Cond. Mat. {\bf24}, 284129 (2012).

\bibitem{romanczuk}
P. Romanczuk, M. B\"ar, W. Ebeling, B. Lindner, and L. Schimansky-Geier, Eur. Phys. J. Special Topics {\bf 202}, 1 (2012).

\bibitem{zoettl}
A. Z\"ottl and H. Stark, Phys. Rev. Lett. {\bf 108}, 218104 (2012).

\bibitem{Cates2012}
M. Cates, Rep. Prog. Phys. {\bf75}, 042601 (2012).

\bibitem{elgeti2015}
J. Elgeti, R. G. Winkler, G. Gompper, Rep. Prog. Phys. {\bf78}, 056601 (2015).

\bibitem{Polin}
M. Polin, I. Tuval, K. Drescher, J. P. Gollub, R. E. Goldstein, Science {\bf325}, 487-490 (2009).

\bibitem{Schwarz}
J. Schwarz-Linek, J. Arlt, A. Jepson, A. Dawson, T. Vissers, D. Miroli, T. Pilizota, V. A. Martinez, W. C. K. Poon. Colloids Surf. B {\bf137} 2-16 (2016).

\bibitem{Theurkauff2012}
I. Theurkauff, C. Cottin-Bizonne, J. Palacci, C. Ybert, L. Bocquet. Phys. Rev. Lett. {\bf108}, 268303 (2012).


\bibitem{Palacci2013}
J. Palacci, S. Sacanna, A. Preska-Steinberg, D. J. Pine, P. M. Chaikin. Science {\bf339}, 936940 (2013).

\bibitem{Buttinoni2013}
I. Buttinoni, J. Bialke, F. K\"ummel, H. L\"owen, C. Bechinger, T. Speck. Phys. Rev. Lett. {\bf110}, 238301  (2013).

\bibitem{Liebchen}
B. Liebchen, D. Marenduzzo, I. Pagonabarraga, M. E. Cates. Phys. Rev. Lett. {\bf115}, 258301 (2015).

\bibitem{Lozano2016}
C. Lozano, B. ten Hagen, H. L\"owen, C. Bechinger, Nature Comm. {\bf 7}, 12828 (2016).

\bibitem{Borge2014}
B. ten Hagen, F. K\"ummel, R. Wittkowski, D. Takagi, H. L\"owen, and C. Bechinger, Nature Comm. {\bf 5}, 4829 (2014).

\bibitem{Palacci2016}
J. Palacci, A. Abramian, S. Sacanna, J. Barral, K. Hanson, A. Y. Grosberg, D. Pine, and P. M. Chaikin, Science Advances {\bf 1}, e1400214 (2015).

\bibitem{BergRT}
H. C. Berg, D. A. Brown, Nature {\bf239}, 500-504 (1972).

\bibitem{Rupprecht}
J. F. Rupprecht, N. Waisbord, C. Ybert, C. Cottin-Bizonne, L. Bocquet, Physical review letters {\bf116} (16), 168101 (2016).


\bibitem{catesRT}
M. E. Cates and J. Tailleur. EPL (Europhysics Letters), {\bf101}, 2 (2013).

\bibitem{RubenNJP}
J. R. Gomez-Solano, C. Bechinger, New J. Physics {\bf17}, 103032 (2015).

\bibitem{Li}
G. Li, A. M. Ardekani, Physical review letters {\bf117} (5), 118001 (2016).

\bibitem{Yazdi}
S. Yazdi, A. M. Ardekani, A. Borhan Physical review E {\bf90}, 043002 (2014).

\bibitem{RubenPRL}
J. R. Gomez-Solano,  A. Blokhuis, C. Bechinger, Phys. Rev. Lett. {\bf116}, 138301 (2016).

\bibitem{gomezsolano1}
J. R. Gomez-Solano and C. Bechinger,  EPL {\bf 108} 54008 (2014).

\bibitem{Bird}
R. B. Bird, R. C. Armstrong, O. Hassager, Dynamics of polymeric liquids â€” Vol. 1: Fluid Mechanics  Vol. {\bf1}  (1987).

\bibitem{das}
S. Das, A. Garg, A. I. Campbell, J. Howse, A. Sen, D. Velegol, R. Golestanian, and S. J. Ebbens, Nature Comm. {\bf 6}, 8999 (2015).

\bibitem{popescu}
W. E. Uspal, M. N. Popescu,  S. Dietrich  and  M. Tasinkevych, Soft Matter {\bf 11}, 434-438 (2015).

\bibitem{brenner}
H. Brenner, Advances in Chemical Engineering  {\bf 6},287-438 (1966).

\bibitem{samin}
S. Samin and R. van Roij, Phys. Rev. Lett. {\bf 115}, 188305 (2015).


\bibitem{rubenSPP}
J. R. Gomez-Solano, S. Samin, C. Lozano, P. Ruedas-Batuecas, R. van Roij, C. Bechinger, arXiv:1709.06339.

\bibitem{hartmut}
S. Babel, B. ten Hagen and H. L\"owen, J. Stat. Mech. P02011 (2014) .

\bibitem{footnote1}
 The particle's orientation dynamics  is much faster than what is expected for a Newtonian liquid with the same viscosity $\eta_0$  \cite{RubenPRL} where an enhancement of the rotational diffusion up to a factor of 250 has been observed. 

\bibitem{zia}
R. N. Zia and J. F. Brady, J. Rheol. {\bf 57}, 457 (2013).

\bibitem{footnote2}
In \cite{RubenNJP} have shown, that the elastic recoil proceeds via a double-exponential decay. However, for small velocities (below 1 $\mu m/s$), the amplitude of the shorter relaxation time can be neglected.



\end{thebibliography}
\end{document}